\documentclass{article}

\usepackage{arxiv}

\usepackage[utf8]{inputenc} 
\usepackage[T1]{fontenc}    
\usepackage{hyperref}       
\usepackage{url}            
\usepackage{booktabs}       
\usepackage{amsfonts}       
\usepackage{nicefrac}       
\usepackage{microtype}      
\usepackage{lipsum}

\usepackage[utf8]{inputenc} 
\usepackage[T1]{fontenc}    
\usepackage{hyperref}       
\usepackage{url}            
\usepackage{booktabs}       
\usepackage{amsfonts}       
\usepackage{nicefrac}       
\usepackage{microtype}      
\usepackage{lipsum}
\usepackage{times}
\usepackage{soul}
\usepackage{url}
\usepackage{graphicx}
\usepackage{amsmath}
\usepackage{bbold}
\usepackage{booktabs}
\usepackage{graphicx}
\usepackage{amsmath}
\usepackage{caption}
\usepackage{subcaption}
\usepackage{float}
\usepackage{array}
\usepackage{microtype}
\usepackage[round]{natbib}
\usepackage{bbm}
\usepackage{color}
\usepackage{dsfont}
\usepackage{tikz}
\usepackage{multirow}
\def\checkmark{\tikz\fill[scale=0.4](0,.35) -- (.25,0) -- (1,.7) -- (.25,.15) -- cycle;}

\title{Mechanical TA 2: A System for Peer Grading with TA Support }

\author{
  Hedayat Zarkoob\\
  Department of Computer Science\\
  University of British Columbia\\
  Vancouver, Canada\\
  \texttt{hzarkoob@cs.ubc.ca} \\
   \And
 Farzad Abdolhosseini \\
  Department of Computer Science\\
  University of British Columbia\\
  Vancouver, Canada \\
  \texttt{farzadab@cs.ubc.ca} \\
  \And
   Kevin Leyton-Brown \\
  Department of Computer Science\\
  University of British Columbia\\
  Vancouver, Canada \\
  \texttt{kevinlb@cs.ubc.ca}\\
}

\begin{document}
\maketitle
\begin{abstract}

\emph{Mechanical TA 2} (MTA2) is an open source web-based peer grading application that leverages trusted TA graders to incentivize high-quality peer review. A previous, prototype implementation of MTA proved the value of the concept, but was neither suitable for use at scale nor easily extensible; MTA2 is a complete reimplementation of the system that overcomes these hurdles. MTA2 serves two, interconnected purposes: facilitating practical peer grading and serving as a testbed for experimentation with different peer grading mechanisms. The system is characterized by a modular design that makes customization easy; support for dividing students into different pools based on their peer-grading prowess; mechanisms for automated calibration and spot checking; and the ability for students to appeal grades and to give feedback about individual reviews. 
\end{abstract}



\section{Introduction}
\label{sec:intro}


\newcommand{\minisec}[1]{\vspace{.4em}\noindent\textbf{#1.~}}

Peer grading can improve students' learning of course material by encouraging critical thinking about the work of their classmates and can also help with the delivery of large classes by reducing grading work for teaching assistants (TAs) and instructors~\citep{sadler2006impact}. 
A wide variety of different peer grading systems already exist~\citep{de2014crowdgrader,Calibratedpeerreview,Peergrade,gehringer2001electronic,pare2008peering}; we survey them in Section~\ref{sec:relatedwork}. All of these systems either work without relying on TAs at all or use TAs only at a very basic level (e.g., resolving student appeals). This is appropriate for courses with extremely inadequate numbers of TAs, such as free massive, open, online courses (MOOCs)~\citep{kulkarni2013peer}. But when more TAs are available (e.g., in many university classes with a few hundred students), they can help to overcome three key drawbacks with current peer grading systems, which we dub \textit{incentives}, \textit{standardization}, and \textit{leverage}.

\minisec{Incentives} Peer grading systems only work when students take their grading tasks seriously. While some students behave altruistically or recognize that peer grading helps their own learning, others are motivated by more immediate incentives, e.g., save time by not putting in the effort to grade others.
Many peer grading systems allow students to appeal when they disagree with a peer review \citep{de2014crowdgrader,Peergrade}; this has the effect of discouraging reviewers from giving low grades. Some systems further reward students for choosing grades that are similar to those assigned by other peers to the same submissions~\citep{de2014crowdgrader,hamer2005method}. Unfortunately, neither of these mechanisms discourages students from giving very high grades to every single submission~\citep{de2016incentives,gao2016incentivizing}, offering students a low-effort alternative to grading sincerely. Experimental evidence shows that human subjects do respond to these incentives \citep{gao2014trick}; we have also observed such behavior in our own classroom teaching.

TAs can overcome this problem by ``spot checking'' a randomly chosen fraction of submissions  \citep{jurca2005enforcing,wright2015mechanical,gao2016incentivizing,zarkoob2019report}. When a submission is spot checked, a TA grades the submission carefully, evaluates the reviews given to that submission, and rewards or punishes the graders (based on the quality of the feedback they offered, the distance between their assigned grades and the TA grades, or other factors). When spot checking is performed to a sufficient degree, students stand to lose more than they stand to gain from dishonest grading, and hence they have strong incentives to grade sincerely \citep{de2016incentives,gao2016incentivizing}. 


\minisec{Standardization} Peer grading systems should teach students to grade consistently and accurately. Otherwise, different students are likely to assign different grades to the same work~\citep{Calibratedpeerreview,wright2015mechanical,wang2019your,kulkarni2013peer}. There are two key ways that TAs can help in this task. The first is by preparing so-called \emph{calibration submissions}, such as submissions from previous years, which can then be assigned to all students in the class \citep{Calibratedpeerreview}. Immediately after a student grades a calibration submission, they can be shown the TA grade and given its rationale; such instantaneous feedback has been shown to aid learning \citep{kulik1988timing}. The second is to let TAs give personalized feedback to students on their grading performance. Calibration can work with very limited TA resources, but of course it is also valuable to have TAs provide direct feedback on actual peer grading (triggered, e.g., by a student appeal, a random spot check, or some other signal such as a pattern of suspiciously similar grades). 

\minisec{Leverage} Peer grading systems must find a way to rely on students who are able to grade their peers reliably---and hence need a relatively low level of supervision---without being derailed by others who are less skilled or more erratic peer graders. In order for a peer grading system to scale, it must trust students who have demonstrated competence. However, students who struggle to grade accurately can lower the average quality of peer grades and undermine student acceptance of the system. TAs can be useful both for identifying such students and for helping to teach them how to grade more effectively.


\emph{Mechanical TA} (MTA) is a system that facilitates peer grading with TA support; it provides all three of the properties just discussed (incentives, standardization, and leverage). We survey key properties of the system in Section 3. A prototype implementation of MTA (MTA1) was used in a computer science course called ``Computers and Society'' for several years. A study published at SIG-CSE \citep{wright2015mechanical} demonstrated the system's usefulness: specifically, MTA1 significantly reduced the need for TA labor in a large class and its use of spot checking, calibration, and division of students into multiple pools helped students to grade more accurately and to learn the course material. However, MTA1 was highly specialized to the Computers and Society course and many of its design choices were hard coded, making the system difficult to change, which limited its usefulness in other settings. 




MTA2 is a complete reimplementation of MTA that improves performance, reliability, and extensibility and also introduces various new features. The design of MTA2 has modularity at its core, allowing 
instructors to choose the features they want based on the workload of their courses and making it easy to change system logic like the rules according to which students move between pools. As already indicated, the design of peer grading mechanisms is an active area of academic research; MTA2's modular design also makes it a good environment for evaluating such mechanisms (e.g., different spot checking or grade aggregation mechanisms). 
To our knowledge, MTA2 is the first peer grading application that offers this level of flexibility. 


Beyond its architectural advantages, MTA2 also offers four key new features that make it applicable in a much wider range of contexts than MTA1. First, MTA2 allows students to submit assignments as ASCII text, PDFs, or answers to multiple choice questions~\footnote{MTA1 only supported plain text.}. 
When text, PDF, or code files are uploaded, the result is rendered directly in the browser (e.g., for grading) rather than needing to be downloaded. Second, students can flag inappropriate reviews independently of the appeal process. This is important because it is a best practice to aggregate peer grades in a way that reduces sensitivity to outliers \citep{hamer2005method}, e.g. by taking the median of peer grades. While this makes grading fairer, it also means that students may not be motivated to appeal when they receive a single inappropriate review. In our early deployment of MTA2, we found that this change can lead to a considerable increase in student satisfaction. 
Third, MTA2 allows instructors to upload groups of student submissions as a single zip file. This allows the ingestion of exams and quizzes conducted outside of MTA, allowing them to be peer graded as well. 
Finally, MTA2 offers advanced SAML2 integration, allowing users to log in using  organizational accounts (e.g., university-wide login systems), increasing ease of use, security, and regulatory compliance.

The rest of this paper is organized as follows. Section~\ref{sec:relatedwork} describes existing peer grading systems and explains how they differ from MTA2. Section~\ref{sec:features} introduces the main features of MTA2 and Section~\ref{sec:userinterface} describes the main components of the system's user interface. 
Section~\ref{sec:evaluation} describes deployment of MTA2 and Section~\ref{sec:conclusion} concludes and discusses future directions.

\section{Related Work} 
\label{sec:relatedwork}
To our knowledge, Calibrated Peer Review (CPR)~\cite{Calibratedpeerreview}, Crowdgrader~\citep{de2014crowdgrader}, Peer Grader (PG)~\citep{gehringer2001electronic}, PeerScholar~\cite{pare2008peering}, and Peergrade~\citep{Peergrade} are the most commonly used peer grading applications.  Tables~\ref{table:comparison} and~\ref{table:featurecomparison} give an overview of MTA2, MTA1, and each of these applications, comparing them in terms of incentives, standardization, and leverage, and also in terms of several desirable system properties. 

Calibrated Peer Review (CPR)~\cite{Calibratedpeerreview} is designed for reviewing writing assignments. CPR pioneered the use of calibration submissions to help students standardize their grading. The system requires students to perform such calibration before grading each assignment. MTA2 largely follows CPR in its use of calibration, except that it allows students who have demonstrated grading proficiency to stop performing calibration.
CPR uses a consensus grade for each assignment to give students some incentive for grading accurately, though it does not offer a mechanism for discouraging all students from reporting high grades.
 
CrowdGrader~\citep{crowdgrader_2013,de2014crowdgrader,crowdgrader_2016} and PeerGrade~\citep{Peergrade} are two commercial peer grading web application and provides many useful features to the instructor~(see Table~\ref{table:featurecomparison}). They have been built to be fast, simple, and scalable to large class sizes. Similar to CPR, CrowdGrader gives students partial incentives via a consensus grade mechanism. PeerGrade aims to incentivize accurate grading by asking reviewers to evaluate each other's reviews. (Note that this does not provide stronger incentives in a game theoretic sense; as before, students can avoid performing the effort of sincere reviewing if they all simply give each other high grades.) 


PG~\citep{gehringer2001electronic} is another peer grading application designed and used for computer science courses. A student in PG can choose either to submit new work or to review the work of other people. PG enables reviewers to communicate with authors during an initial feedback phase, after which authors can revise their submissions; finally, reviewers assign a grade. PG aims to encourage accurate grading by having authors evaluate reviewers. It records the best student submission, allowing it to be used for educational purposes later.  
  
PeerScholar~\citep{pare2008peering} is another application which focuses on implementing peer grading in large classes. Similar to MTA, it started by supporting writing assignments, but, its later versions included different assignment formats. The early version of PeerScholar asked students to write separate essays demonstrating comprehension of an assigned reading and reacting to it; these essays are then peer graded.


Several further peer-reviewing systems focus on gathering fast feedback on draft versions of a student assignment \citep{cho2007scaffolded,kulkarni2015peerstudio,politz2014captainteach,mulder2007praze}, rather than using these systems to assign final grades. 
Notably, CaptainTech~\citep{politz2014captainteach} used the idea of spot-checking on seeded solutions in a programming course, but in a much more limited way than MTA. Its designers argue that since their system does not use student reviews to calculate final grades, they only lightly used spot checking to check students' feedback quality. 

We are also aware of two more recent peer grading applications: Kritik~\citep{Kritik} and Peerceptiv~\citep{peerceptiv}. Kritik is an AI based peer grading application that focuses on finding fair, accurate, and high quality grades. Peerceptiv is developed based on~\cite{cho2007scaffolded}'s design with data driven algorithms to assign final grades to students.
\begin{table*}[h]
\centering
 \begin{tabular}{p{1in}p{0.9in} c c c c c c c c} 
 \toprule
 \textbf{Property} & \textbf{Method} &\textbf{PeerScholar} & \textbf{PG} & \textbf{CPR} & \textbf{CrowdGrader} & \textbf{PeerGrade} & \textbf{MTA1} & \textbf{MTA2} \\  
\midrule
\multirow{3}{*}{\emph{Incentives } via}& TA spot checks & -- & -- & -- & -- & -- & \checkmark & \checkmark \\
& grade consensus & -- & -- & \checkmark & \checkmark & \checkmark & -- & \checkmark\\
& review the reviewer & \checkmark & \checkmark & -- & -- & \checkmark & -- & \checkmark \\
\hline
\multirow{2}{*}{\emph{Standardization} via}  & calibration submissions & --  & -- & \checkmark & -- & -- & \checkmark & \checkmark \\
& instructor/TA feedback & --  & -- & \checkmark & \checkmark & \checkmark & -- & \checkmark \\
\hline
\emph{Leverage} via  &dividing graders into pools & -- & -- & -- & -- & -- & \checkmark & \checkmark \\
 \bottomrule
\end{tabular}
\caption{Comparing peer grading applications: Our key desiderata.}
\label{table:comparison}
\end{table*}

\begin{table*}[h]
\centering
 \begin{tabular}{p{1.9in}  c c c c c c c} 
 \toprule
 \textbf{Feature} &\textbf{PeerScholar} & \textbf{PG} & \textbf{CPR} & \textbf{CrowdGrader} & \textbf{PeerGrade} & \textbf{MTA1} & \textbf{MTA2} \\  
\hline
Instructors can enable and disable features  & \checkmark  & -- & \checkmark & -- & \checkmark & -- & \checkmark \\ 
Instructors can change system logic  & -- & -- & -- & -- & -- & -- & \checkmark \\ 
Support for different assignment types & \checkmark  & \checkmark & -- &\checkmark & \checkmark & -- & \checkmark \\
Students can give feedback about individual reviews  & -- & -- & -- & -- & \checkmark & -- & \checkmark \\
Provide data-driven insights to instructors & -- & -- & -- & -- & \checkmark & -- & -- \\
Students can hand in group assignments  & -- & -- & -- & \checkmark & \checkmark & -- & -- \\
Support for checking submission similarity & -- & -- & -- & \checkmark & -- & -- & -- \\
Students and instructors can discuss submissions & \checkmark & \checkmark & -- & \checkmark & -- & -- & -- \\
Instructors can choose to grade individual submissions & -- & -- & -- & \checkmark & \checkmark & -- & \checkmark \\
Easy integration with SAML2 & -- & -- & -- & -- & -- & -- & \checkmark \\
Students can review themselves (self-assessment) & -- & -- & \checkmark & -- & \checkmark & -- & \checkmark  \\
 \bottomrule
\end{tabular}
\caption{Comparing peer grading applications: Other major features.}
\label{table:featurecomparison}
\end{table*}

\section{MTA2: Key Features}
\label{sec:features}


In this section, we describe MTA2's key features. We explain how these deliver the properties discussed in the introduction (incentives, standardization, and leverage) and help MTA2 to be a high-performance, reliable, and extensible peer grading application.  


MTA2 makes it possible for instructors to toggle any of the following features: supervised and independent pools, calibration, spot-checking, giving feedback about individual reviews, and integration with SAML2; we will go on to describe each of these features in turn. For users requiring finer grained customization, the MTA2 codebase is written in Python, which makes modifying its behavior simple. The source code is divided into eight directories based on function: assignment, review, evaluation, calibration, grade, account, course, and homepage. The contents of each of these directories is further decomposed into three parts: \textit{templates and views}, which control the UI; \textit{base}, which contains the logic; and \textit{models}, which contain the definitions of data types. 
Hence a newcomer attempting to modify a specific behavior can quickly identify the part of source code that requires modification. For example, by default, MTA2 calculates final grades by taking the median of the grades given to each submission. However, an instructor might prefer a more complicated grade aggregation algorithm, such as Olympian Average. To modify the grade calculation logic one needs only to modify the \texttt{base.py} script inside the \texttt{peer\_grade} directory.

\subsection{Incentives}
MTA2 uses four main features to provide students with sufficient incentives for accurate grading: spot checks, appealing, giving feedback about individual reviews, and using the consensus grade .
    
\subsubsection{Spot checking}
In MTA2, a spot check has two parts. The first is a \emph{TA review} of the assignment: the chosen assignment is graded by a TA, using the same grading rubric used by students. Second is \emph{TA evaluation} of the peer graders: the TA uses a separate rubric to evaluate the reviews submitted by students who graded the spot checked assignment, giving a grade on a 10-point scale. 
By default, the system lets the instructor choose the number of assignments that need to get spot checked and assigns TAs to students uniformly at random. However, this logic can be revised easily, for example targeting high grades. 
Furthermore, TAs are free to browse through assignments and conduct additional spot checks, based e.g. on reviewer identity, degree of agreement between reviewers, or peer grade assigned.
\subsubsection{Appealing}

Students can appeal when they disagree with a grade assigned by peer reviewers. If the appeal provides a convincing argument, instructors or TAs regrade the assignment and also evaluate every peer review that the assignment has received.

\subsubsection{Giving feedback about individual reviews}
MTA2 also allows students to provide feedback about individual reviews. The procedure is similar to appealing, but it is supported via a separate mechanism for two reasons. First, we found it useful to allow students to identify rude, unfair, or low quality reviews even when these reviews do not affect the student's overall grade. If a student believes that a review is inappropriate, they can flag the review and provide their reasoning. The report is then assigned to a TA as in an appeal. If the TA confirms that the report is inappropriate it no longer affects the assignment's grade. The reviewer is also notified and the TA may take additional action outside the MTA2 system. Second, we wanted to allow students to give positive feedback about particularly good reviews, to help reward students for making exceptional efforts.

\subsubsection{Using the consensus grade}

When an assignment is not spot checked, instructors can choose to reward graders by measuring how close their grades are to the consensus grade. We note, however, that recent work has shown that the inclusion of such a mechanism \emph{increases} the amount of spot checking required to provide strong incentives to graders~\citep{gao2016incentivizing}; thus, it is not enabled in MTA2 by default.


\subsection{Standardization} \label{subsec:calibration}
There are two main ways that MTA2 facilitates standardization.

\subsubsection{Calibration submissions}

Calibration submissions have known ``ground truth'' grades; for example, they might be carefully graded submissions from previous offerings of a course. They are graded using the same rubric as student submissions and can either be presented separately to students in the UI (to support a workflow like requiring a calibration review before peer reviews) or mixed together with student submissions (to drive TA spot checks by detecting students who grade poorly).
MTA2 auto-grades calibration reviews by starting with 10 points and then, for each rubric question, subtracting the square of the difference between the student grade and the ground truth grade. For example, if there are two rubric questions and a review is off by one for two of the questions, the score is $10-1^2-1^2=8$. If the review is off by two for only one question, the score is $10-2^2=6$. This is motivated by the idea that a student who understands how to grade well might make small errors, while bigger errors reflect more fundamental misunderstandings. Of course, this formula can be easily changed.

\subsubsection{TA and instructor feedback}

TAs and instructors have the opportunity to give student graders feedback about their reviews whenever a submission is spot checked or appealed. This personalized feedback provides another avenue for students to standardize their understanding of the grading rubric. 

\subsection{Leverage: Supervised/independent pools} \label{subsec:pools}

MTA is able to maintain different \emph{pools} of students, where students are only assigned peer reviews for others in the same pool.
The first, ``supervised'' pool guarantees that every submission and review submitted by students will also be graded by a TA, with only the TA grade counting for evaluation. The purpose of this pool is to ensure that students do not assign binding peer grades until they have shown themselves to be fair and consistent graders. At the beginning of the course, all students are assigned to the supervised pool. They graduate to the second, ``independent'' pool by demonstrating consistent performance either in calibration reviewing or in real peer review in the supervised pool. (Recall that both calibration reviews and peer-grading reviews are assessed on 10-point scales. By default, MTA promotes a student to the independent pool once they obtain at least 35 points over their five most recent peer-grading or calibration reviews.) Students in the independent pool are evaluated for the quality of their grading only if they are spot checked, the submission they have graded is appealed, or their review is flagged as inappropriate.


The use of supervised and independent pools offers a second benefit for calibration: they dramatically reduce TA grading when a new course starts. Performing calibration reviews can help supervised students to become independent more quickly---often, even before submitting the first assignment. Instructors can also require supervised students to perform additional, weekly calibration reviews, both to reduce TA workload and to help weaker students to learn to peer grade effectively.



\subsection{Other major features}
MTA2 also has a variety of new features that do not directly address incentives, standardization or leverage. Here we describe a two of the most important.




\subsubsection{Different assignment types}
In MTA2, an assignment can be created by specifying the assignment's name, type, problem statement, release date, and deadline.
Instructors and TAs can also set a late unit period for an assignment, which is the number of days the students' submissions are allowed to be late. MTA2 keeps track of the total number of late days each student has accumulated across the whole course. Once this number reaches the limit set by the instructor at the beginning of the course, the system stops accepting late submissions. In addition to the late unit period, an assignment can also have a grace period, which is a short amount of time after the deadline during which the submission will not be considered late. 
MTA2 supports three types of assignments.

\minisec{Text} Students submit answers by typing text directly into a text box. Instructors can set a limit for the number of characters or words that can be submitted.

\minisec{Upload} Assignments can also be configured to solicit the upload of PDF, text, or code files. Uploaded files are rendered in the browser directly. When code is submitted, MTA2 highlights syntax automatically. Instructors can also upload a group of student submissions as a zip file, with each student's submission named as their MTA username. This feature lets instructors use the grading features of MTA2 for exams and assignments that happen outside the MTA2 framework.

\minisec{Quiz} Assignments can be multiple-choice quizzes, which can be graded automatically. The system can be configured to show students only final grades (not correct answers) after submission. Instructors can make a quiz a requirement for doing other types of assignments, which is helpful for making sure that students satisfied a prerequisite (e.g., did a reading) before submitting. The instructor can limit the number of times a quiz can be submitted.


\subsubsection{Easy integration with SAML2}
Security Assertion Markup Language 2.0 (SAML 2.0) defines standard protocols for the exchange of authentication and authorization between domains. It can be used to provide cross-domain single sign on. In many universities and different organizations, users have a universal account that can be used to log into different applications provided by that organization. MTA2 can be integrated into such systems that use the SAML2 standards and can therefore delegate the administration of user accounts to the central organization.
%
%
Internally, we use the \texttt{django-saml2-pro-auth} package which is a wrapper for the \texttt{python-saml} package.

\section{MTA2: User Interface}
\label{sec:userinterface}
    

\begin{figure*}
\centering
\begin{subfigure}[t]{.485\textwidth}
    \vskip 0pt  
    \centering
    \includegraphics[width=1\linewidth]{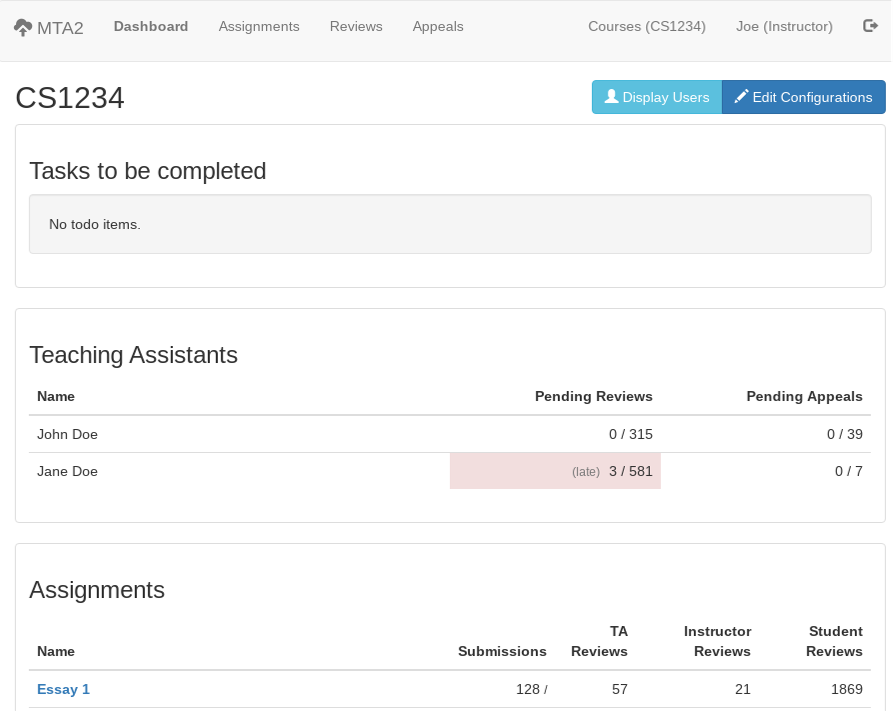}
\end{subfigure}
\begin{subfigure}[t]{.51\textwidth}
    \vskip 0pt  
    \centering
    \includegraphics[width=1\linewidth]{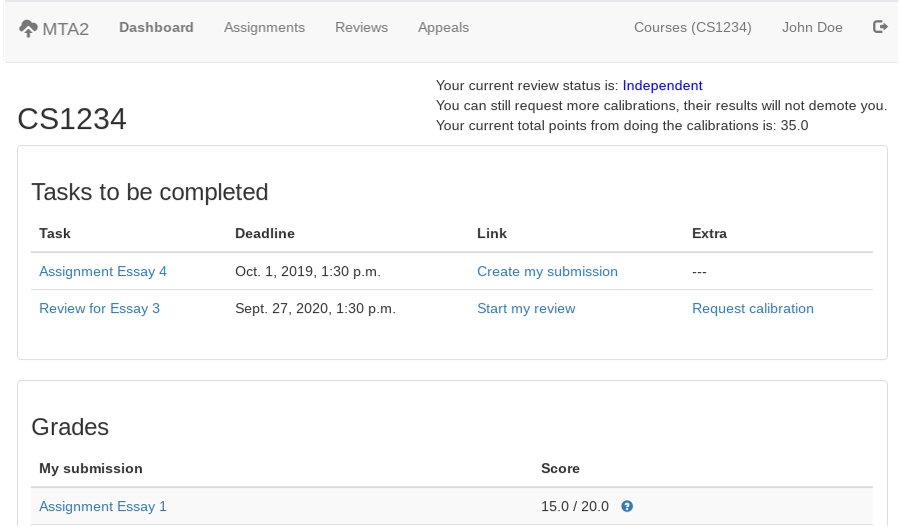}
\end{subfigure}
\caption{The dashboard tab for the instructor (on the left) and the students (on the right).}
\label{fig:dashboard_tab}
\end{figure*}

%
When a user logs into MTA2, they see a list of the courses they are associated with and the role they have in each. Users can associate themselves with a new course by entering a given code.
Upon choosing a course, every user sees an interface divided into four main tabs: dashboard, assignment, review, and appeal. 

\subsection{Dashboard}
The dashboard is a one-screen overview shown at login; for instructors, it shows assigned tasks, a summary of course statistics, and a link to a list of all students~(see Figure~\ref{fig:dashboard_tab} on the left). By clicking on a student's name in this list, TAs and instructors can see every task that the chosen student has done throughout the course and the corresponding grades they were given. Finally, the course configuration (course name, visibility to students, enrollment codes, etc) can be viewed; which can only be edited by the instructors.

For students, the dashboard contains two main tables, as illustrated in Figure~\ref{fig:dashboard_tab} on the right. The first  shows  assigned tasks and deadlines. The second shows grades for previous assignments. Students can view their submissions and their reviews in detail by clicking on the associated grade. They can also appeal or flag a review if necessary. Student reviews are only available after the student review deadline is passed. The final score of each submission is shown once the TA review deadline is reached. (Observe that TAs might spot check an assignment and thereby change the assigned grade.) When multiple pools are enabled, the dashboard also indicates whether the student is supervised or independent.


\subsection{The Assignment tab}

Instructors and TAs can see the list of all assignments and create new ones using this tab. They can also view the submissions for each assignment by displaying the submission list. Instructors can furthermore determine whether an assignment is visible to students and set deadlines. 

For students, the Assignment tab shows students the list of their assignments; for assignments that are not yet due, students can edit their submissions. Students can also see and appeal their grades in this tab once the deadline for TA reviews has passed. 

\subsection{The Reviews tab}

Reviewing an assignment means reading a submitted piece of work and then completing a rubric. 
Rubric questions ask graders to evaluate an assignment in a particular dimension by choosing the most appropriate among a set of multiple choices. Each rubric question can also have a reasoning field, which asks the grader to provide a rationale for their choice. The rationale field can be set to require entries between a minimum and maximum length.

Submitted reviews can later be evaluated by TAs (using a different rubric). Instructors and TAs can assign reviews and evaluations, create rubrics, and see review stats from the review tab.
If an assignment has more than one question, it is possible to assign different TAs to grade different questions. The algorithm used to assign TAs to assignments or to assign spot checks can be changed straightforwardly.

Instructors and TAs can also access the list of submitted reviews and view each one from this tab. 
When viewing a student review, instructors and TAs can request an evaluation for the review. Note that review evaluation is most naturally part of the spot-checking process. However, if instructors or TAs encounter a review that requires evaluation (e.g., via an appeal), they can also do so directly.

For students, the reviews tab displays two tables: one that shows pending reviews that the student still needs to submit and another that shows reviews that the student has previously submitted. The students can still edit their submitted reviews until the review deadline.

\subsection{The Appeals tab}
\label{subsec:appeals}
The appeals tab shows each TA the appeals that are assigned to him or her along with their current status. They can also reassign an appeal that they cannot resolve themselves. Instructors can see all the appeals. For students, the appeals tab keeps track of all previously submitted appeals and their current status. New appeals are submitted via the assignment tab.

\subsection{Admin interface}
Certain instructors or TAs can be designated as administrators, meaning that they are afforded fine-grained control over course data via \href{https://docs.djangoproject.com/en/dev/ref/contrib/admin/}{Django Admin} interface, which is accessed via a fifth tab, ``Administration''. This interface allows for searching and filtering the data, and can be customized with new filters. It can also be used for data entry and manipulation. While this feature is not designed to be used frequently in production, it can be useful for testing or trying out new ideas.


\section{Deploying MTA2}
\label{sec:evaluation}

To date, MTA2 has been deployed in three computer science courses, focused respectively on 
algorithms, programming, and computers and society. We have not run a formal user study to evaluate MTA2's new design, but have informally received considerable positive feedback from students and instructors who used the system. In the case of the computers and society course, we have access to some data about the deployment that is worth summarizing here. 

\begin{figure}[t]
	\centering
	\includegraphics[scale=0.46 ]{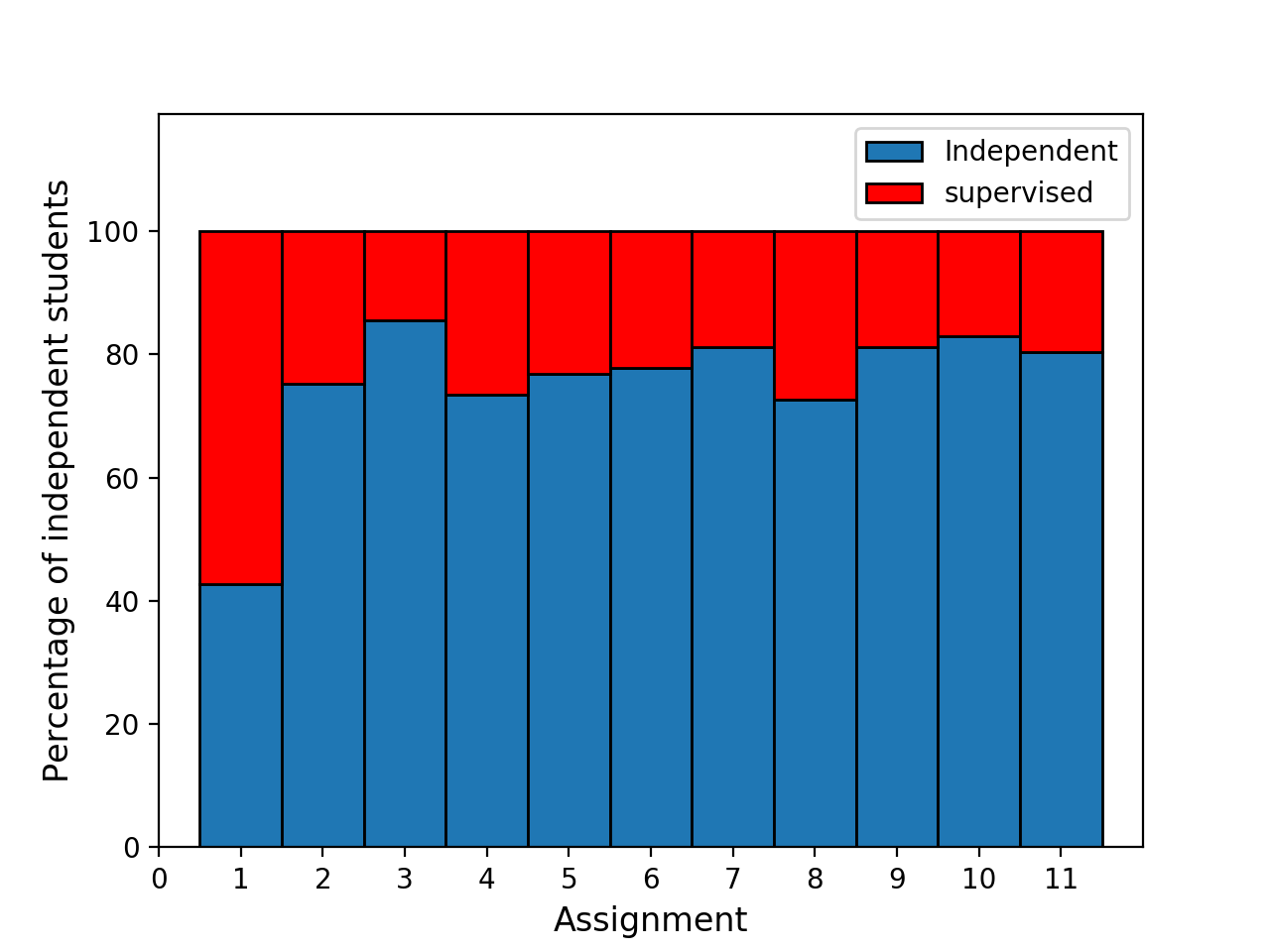}
	\caption{The percentage of independent students as a function of assignment number}
	\label{fig:ind_percentage}
\end{figure}

The computers and society course had 117 students, 11 assignments, 1203 student submissions, 77 calibration essays, 460 TA reviews, and 75 appeals. Figure~\ref{fig:ind_percentage} shows the percentage of students assigned to the independent pool for each assignment~\footnote{For this course, the rule for getting students to the independent pool was slightly more strict than the one described in Section 3.3.}. Observe that by the first assignment, about 40 percent were promoted based on calibration reviews; by the second assignment about 80 percent were promoted, based both on additional calibration and their peer reviews of assignment 1. After this point, the percentage of students in the independent pool fluctuated, due to some students being demoted to the supervised pool. Such demotions were triggered by spot checks, appeals, and reports of individual bad reviews. Overall, these percentages remained around 80 percent, showing that the system succeeded in motivating the vast majority of students to perform high quality peer reviews.

\begin{figure}[t]
	\centering
	\includegraphics[scale=0.46 ]{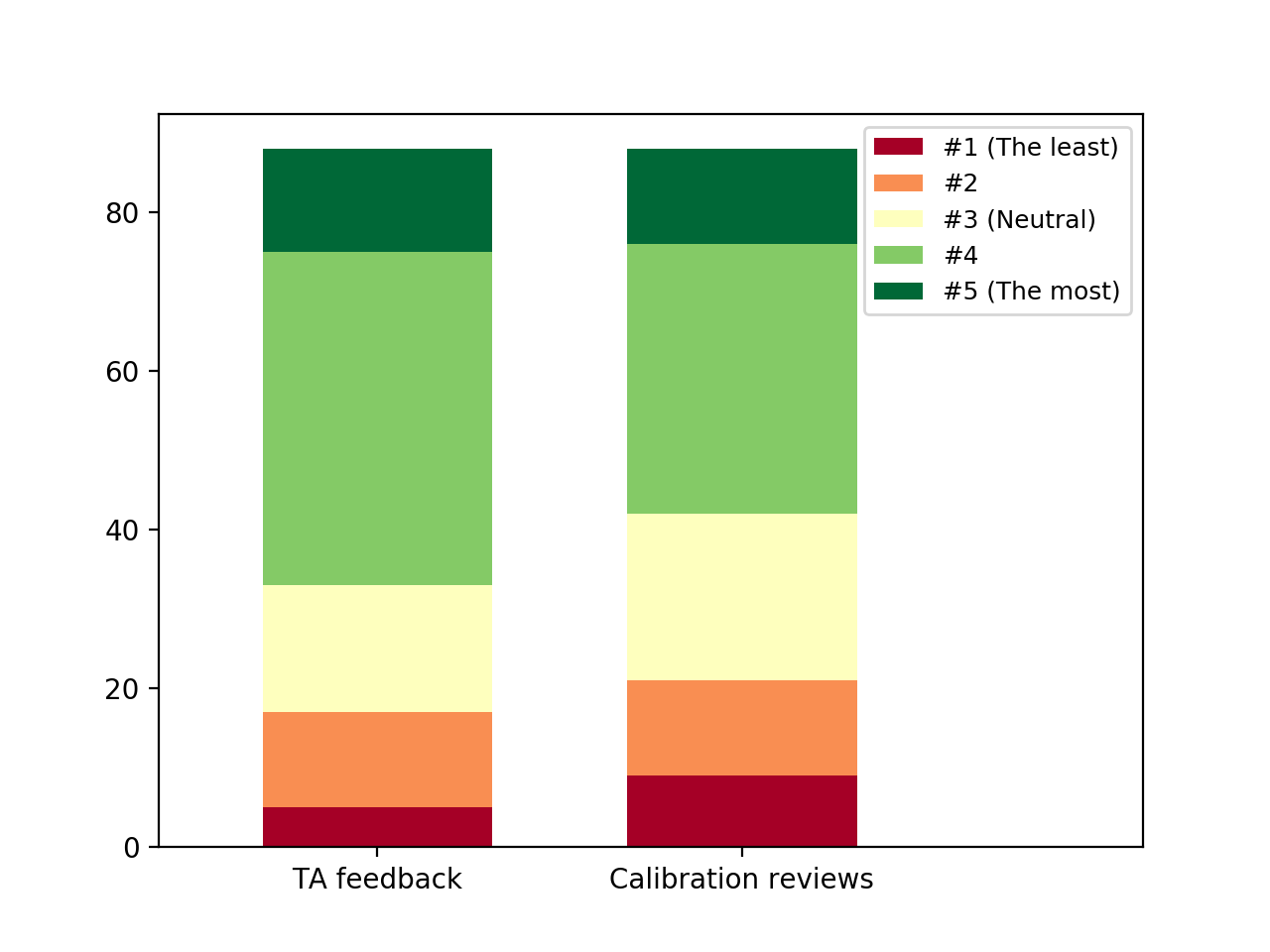}
	\caption{Students' answer to the two questions about how they learned to improve the quality of their peer reviews. The results show that TA feedback and calibration reviews were almost equally useful}
	\label{fig:ta_vs_cal}
\end{figure}

We conducted an end-of-year survey to help us to improve the next course offering. While it was not designed as a research instrument, two categories of responses are useful for investigating the effectiveness of MTA2's design. 
First, two questions asked students about how they learned to improve the quality of their peer reviews. Students reported that TA feedback and calibration reviews were  almost equally effective (see Figure~\ref{fig:ta_vs_cal}); however, note that the distribution of support for TA feedback stochastically dominates that of calibration reviews (i.e., students preferred to receive TA feedback). 

\begin{figure}[t]
	\centering
	\includegraphics[scale=0.46 ]{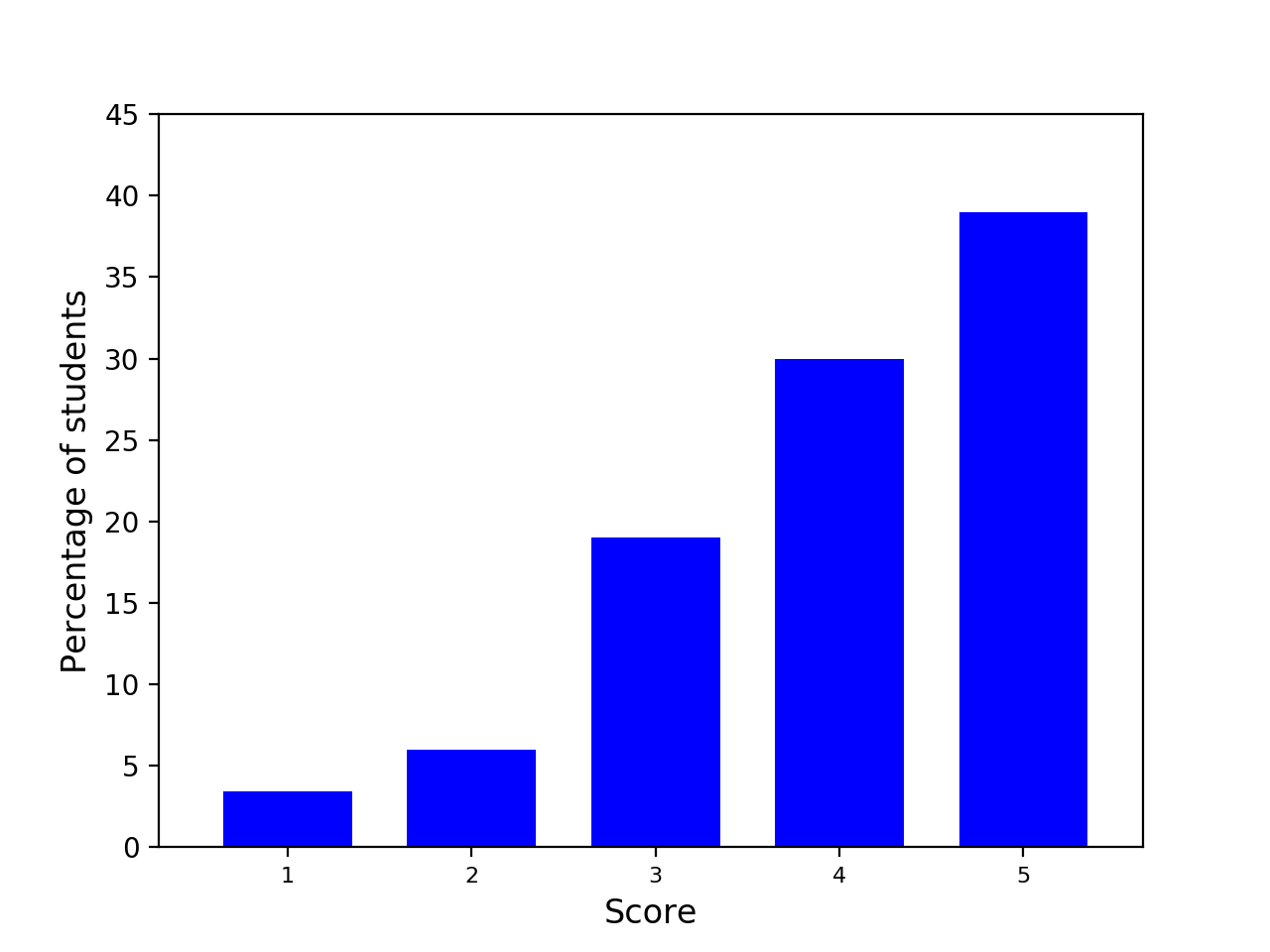}
	\caption{MTA2 user interface ranking}
	\label{fig:user_interface}
\end{figure}

Second, we asked students how user friendly they found the MTA2 interface. Overall, 69\% scored the user interface either 4 or 5 out of 5 (see Figure~\ref{fig:user_interface}). We contrast this with the system's initial implementation (MTA1); while we are aware of no comparable survey data, that system received many oral and written complaints about its user interface, which indeed served as one of the motivations for our system redesign.

\section{Conclusion}
\label{sec:conclusion}
Mechanical TA 2 (MTA2) is an open source, web-based peer grading application that leverages trusted TA graders to overcome three main drawbacks in traditional peer grading: incentives, standardization, and leverage. The system is designed both to facilitate practical peer grading and to support research and experimentation about the design of peer grading mechanisms. MTA2 is characerized by a modular design that makes customization easy; support for dividing students into multiple pools based on their peer-grading ability; mechanisms for automated calibration and spot checking; and the ability for students to appeal grades and to flag inappropriate reviews. 


In future work, we plan to experiment with different peer grading mechanisms using MTA2. We are sure that these experiments will lead us to continued work on the system's user interface and features. 

\bibliographystyle{named}
\bibliography{references}

\end{document}